\documentclass[10pt,conference]{IEEEtran}
\IEEEoverridecommandlockouts
\usepackage[hidelinks]{hyperref}
\usepackage[compress]{cite}
\usepackage{amsmath,amssymb,amsfonts}
\usepackage{algorithmic}
\usepackage{graphicx}
\usepackage{textcomp}
\usepackage{xcolor}
\usepackage{subcaption} % For subfigure support
\usepackage{array}
\usepackage{booktabs}
\usepackage{multirow}
\usepackage{caption}
\usepackage{tcolorbox}
\usepackage{wasysym}
\usepackage{threeparttable} % for table + footnotes
\usepackage{balance}
\usepackage{tikz}
\usepackage{colortbl}
\usepackage[T1]{fontenc}
\usepackage{fontawesome}
\usepackage{adjustbox}
\usepackage{tabularx}
\usepackage{pifont}
\usepackage{enumitem}
\definecolor{deepblue}{HTML}{0c74d4}
\definecolor{lightblue}{HTML}{64abf3}
\definecolor{lightgray}{HTML}{dee0e2}
\definecolor{lightorange}{HTML}{fbb362}
\definecolor{deeporange}{HTML}{e26f1e}

\def\BibTeX{{\rm B\kern-.05em{\sc i\kern-.025em b}\kern-.08em
    T\kern-.1667em\lower.7ex\hbox{E}\kern-.125emX}}
\begin{document}

\newenvironment{leftquote}[1][1em]{%
  \par\noindent
  \begin{list}{}{%
    \setlength{\leftmargin}{#1}% 只设置左缩进
    \setlength{\rightmargin}{0pt}% 右侧不缩进
    \setlength{\parsep}{\parskip}% 保持段间距
  }%
  \item[]%
}{%
  \end{list}\par
}

\title{``\textit{My productivity is boosted, but ...}''  Demystifying Users' Perception on AI Coding Assistants}
\IEEEaftertitletext{\vspace{-1.5em}}  % 调整 -1.5em 以适应你的需求

% \author{\IEEEauthorblockN{Anonymous Authors}}

\author{
\IEEEauthorblockN{Yunbo Lyu\IEEEauthorrefmark{2},
Zhou Yang\IEEEauthorrefmark{3}\IEEEauthorrefmark{1}\thanks{* Corresponding author.},
Jieke Shi\IEEEauthorrefmark{2},
Jianming Chang\IEEEauthorrefmark{4},
Yue Liu\IEEEauthorrefmark{2},
David Lo\IEEEauthorrefmark{2}}
\IEEEauthorblockA{
  \IEEEauthorrefmark{2}Singapore Management University 
  \IEEEauthorrefmark{3}University of Alberta 
  \IEEEauthorrefmark{4}Southeast University}
Email: \{yunbolyu, jiekeshi, liuyue, davidlo\}@smu.edu.sg, zy25@ualberta.ca, jianmingchang@seu.edu.cn
}
% Email: zyang@ncsu.edu
% Email: \{yunbolyu, jiekeshi, yueliu, davidlo\}@smu.edu.sg}
% Email: jianmingchang@seu.edu.cn

\maketitle

\thispagestyle{plain}

\begin{abstract}

This paper aims to explore fundamental questions in the era when AI coding assistants like GitHub Copilot are widely adopted: \textit{what do developers truly value and criticize in AI coding assistants, and what does this reveal about their needs and expectations in real-world software development?}
Unlike previous studies that conduct observational research in controlled and simulated environments, we analyze extensive, first-hand user reviews of AI coding assistants, which capture developers' authentic perspectives and experiences drawn directly from their actual day-to-day work contexts.
We identify 1,085 AI coding assistants from the Visual Studio Code Marketplace. 
Although they only account for 1.64\% of all extensions, we observe a surge in these assistants: over 90\% of them are released within the past two years.
We then manually analyze the user reviews sampled from 32 AI coding assistants that have sufficient installations and reviews to construct a comprehensive taxonomy of user concerns and feedback about these assistants.
We manually annotate each review's attitude when mentioning certain aspects of coding assistants, yielding nuanced insights into user satisfaction and dissatisfaction regarding specific features, concerns, and overall tool performance.
Built on top of the findings—including how users demand not just intelligent suggestions but also context-aware, customizable, and resource-efficient interactions—we propose five practical implications and suggestions to guide the enhancement of AI coding assistants that satisfy user needs.
% Through sentiment analysis, we uncover 12 key findings that reflect what users care about, appreciate, and dislike. 
% Users most value accurate code suggestions, while expressing dissatisfaction with buggy outputs, incomplete responses, and high resource consumption. Dependability emerges as a central concern, especially regarding reliability and trust. 
% Based on our findings, we propose five practical implications to guide the future development of AI coding assistants.

  \end{abstract}

\begin{IEEEkeywords}
    AI Assistant, IDE, User Experience, Programming, Human-Computer Interaction,  Human Factors
\end{IEEEkeywords}

\section{Introduction}
\label{sec:intro}

AI coding assistants,\footnote{In this paper, the terms AI coding assistants, in-IDE AI coding assistants, assistant, and AI extensions are used interchangeably, referring to AI-powered extensions in the IDE, such as GitHub Copilot.} such as GitHub Copilot, have revolutionized software development practices~\cite{mozannar2024reading}.
These assistants leverage Large Language Models (LLMs) to provide code suggestions, ranging from token-level to block-level code completion, bug fixing, testing, and code explanation—covering nearly every aspect of software development~\cite{sergeyuk2025using, chen2024code, yang2025morepair}.
Their ability to accelerate software development and reduce workloads has led to widespread adoption~\cite{liang2024large, github2024copilotstats}.

Although AI coding assistants have seen rapid growth and widespread adoption, recent studies indicate that developers may spend over 50\% of their coding time verifying AI-generated suggestions, which can increase cognitive load and introduce additional validation effort~\cite{mozannar2024reading}.
This highlights the importance of understanding software developers' perceptions and needs.
Such understanding enables assistant developers to better align their tools with actual user requirements and shape the future of software development more effectively.

Recent studies explore developers' perceptions through interviews~\cite{mcnutt2023design,liu2024empirical, prather2023weird}, surveys~\cite{wang2023practitioners, liang2024large}, and user studies~\cite{vaithilingam2022expectation, barke2023grounded, mozannar2024reading}.
However, these studies face limitations, such as small or restricted samples, high costs~\cite{davis2023s}, predefined questions, or a narrow focus on only a few AI coding assistants~\cite{sergeyuk2025human}.

Meanwhile, online resources such as app reviews have proven effective in bridging the gap between developers and users, as they often involve large user bases and extensive feedback~\cite{pagano2013user,martin2016survey,al2019app}.
Yet, few studies have mined reviews for AI coding assistants~\cite{zhang2023demystifying,zhou2025exploring}, and those that do face three notable limitations:
(1) They focus exclusively on GitHub Copilot, overlooking the broader ecosystem of AI coding assistants and missing opportunities for cross-tool comparisons.
(2) Existing taxonomy are built from sources like GitHub Issues and Stack Overflow, which highlight technical problems raised by advanced users~\cite{liao2019status}. 
This limits insight into subjective experiences and positive feedback, failing to capture what users value—not just what they find problematic.
(3) Novice or non-technical users, who may lack the skills or motivation to create GitHub issues, are more likely to leave reviews on the Visual Studio Code Marketplace, where feedback is easier to submit. 
These overlooked voices are essential for understanding real-world perceptions of AI coding assistants~\cite{liao2019status, ford2016paradise}.

An unprecedented opportunity exists to address this research gap: thousands of AI coding assistants are available in the Visual Studio Code (VS Code) marketplace, accompanied by extensive user feedback.
These real-world reviews provide a rich source of information from developers all over the world, offering insights into their needs and expectations.
As the most widely used IDE globally—adopted by 74\% of developers according to Stack Overflow's 2024 survey~\cite{stackoverflow2024survey}—VS Code hosts a diverse range of AI coding assistants~\cite{liu2025protect}.
This rich dataset enables comprehensive analysis of user perceptions across multiple assistants, offering insights at scale and providing robust validation and extension of prior findings.

We first collect 1,085 AI coding assistants identified among 66,053 extensions in the VS Code marketplace~\cite{vscode_marketplace} through a combination of manual and automatic labeling.
We employ GPT-4o to assist in the labeling process and validate its performance on a statistically significant sample, achieving both precision and recall above 96\%.
Our analysis reveals that over 90\% of AI extensions were released in 2023 and 2024, highlighting a recent surge in their adoption.
Additionally, we find that a small number of AI extensions dominate the market, with the top 10 installed extensions accounting for 86\% of total installs.
Moreover, we find that AI coding assistants receive more engagement, as AI extensions receive significantly more feedback and have higher average installs.

The first goal of this paper is to construct a taxonomy of the aspects users discuss regarding AI coding assistants. 
Such a taxonomy helps developers understand what users care about at a high level and which specific characteristics of AI extensions they are particularly concerned with. 
It also supports improvements in development, maintenance, and testing by highlighting key focus areas.
To ensure a systematic taxonomy construction process, we follow best practices for app review analysis~\cite{dkabrowski2022analysing}.
We begin by sampling 361 user comments—statistically representative of the 5,908 reviews collected from 32 AI extensions.
We do not consider extensions with insufficient installations or reviews.
During the coding process, we adopt a bottom-up merging approach following prior work~\cite{humbatova2020taxonomy}.
We first assign one or more labels to each comment, then merge identical labels, group them into subcategories, and finally organize them into top-level categories.
Instead of open card sorting, we use a hybrid card sorting~\cite{conrad2019making}, as the top-level categories are informed by established literature in the mature field of app review mining. 
In our pilot coding of 30 reviews, we initially generated 11 fragmented top-level categories. 
The predefined structure helps ensure conciseness, with pilot labels naturally aligning with the chosen dimensions.
The final taxonomy includes 8 top-level categories (functionality, usability, dependability, supportability, system performance, general experience, comparison, and pricing), 16 subcategories, and 62 distinct labels.

We analyze user feedback on AI coding assistants and identify six key findings. Users value accurate suggestions but often criticize redundant, incomplete, or buggy outputs. 
While many report productivity gains, novices are positive than experienced developers.
Context awareness is a major issue: assistants understand code well but often fail to retain or fetch relevant context. 
Usability also matters—complex onboarding and disruptive interfaces frustrate users.
Response time is generally acceptable, but resource consumption is a common complaint. 
Finally, pricing and ethical concerns affect adoption, with users favoring free tools and questioning monetization of open-source data.

% We further analyze what users care about, like, and dislike regarding AI coding assistants, leading to seven key findings and five practical implications. 
% The most frequently discussed topic is the content of suggestions provided by the assistant. 
% Among these, accurate suggestions receive the most praise—users value instances where the assistant correctly anticipates their coding intentions. In contrast, the most criticized outputs are those that are buggy, redundant, or incomplete.
% Dependability emerges as a major concern, with 78\% of user comments in this category expressing dissatisfaction. 
% Based on this, we suggest that developers prioritize perceived dependability and focus on building user trust.
% Another key finding is that resource consumption is a concern not only when running local models but also with online models, especially when applied to large codebases. 
% Users report performance issues and high resource usage in both scenarios.

The contributions of this paper are:
(1) Identifying 1,085 AI-related extensions from 66,053 VS Code extensions using a hybrid labeling method.
(2) Constructing a three-level taxonomy from 361 user reviews, with 8 categories, 16 subcategories, and 62 labels.
(3) Performing sentiment analysis to uncover six key findings on user preferences and concerns, leading to five practical implications.
(4) Releasing all data, annotation tools, and results in the replication package.

\section{Related Work}
\label{sec:background}

\subsection{Studies on User Perceptions of AI Coding Assistants}
\label{subsec:ai_assisted_programming}

%  define what is ai coding assistant
AI coding assistants are IDE extensions that leverage large language models (LLMs) to support a range of development tasks, including code completion, bug fixing, testing, and code explanation~\cite{sergeyuk2025using}.
These assistants primarily operate through three interaction modes: code completion (ranging from token- to statement-level), chat interfaces, and the emerging agent mode, which can execute terminal commands~\cite{github_copilot_agent_2025}.

Researchers have employed diverse methods to study developers' perceptions of AI coding assistants, including interviews, surveys, user studies, and mining online content~\cite{yang2024robustness}.
Zhou et al.~\cite{zhou2025exploring} mine 476 GitHub issues, 706 discussions, and 142 Stack Overflow posts to build a taxonomy of problems developers encounter with GitHub Copilot.
Zhang et al.~\cite{zhang2023demystifying} analyze similar sources to identify developer practices, challenges, and expectations.

Survey and interview studies are common for capturing user experiences. GitHub’s industry report~\cite{github2024survey} surveys 500 developers on the impact of AI assistants.
Liang et al.~\cite{liang2024large} and Sergeyuk et al.~\cite{sergeyuk2025using} survey 410 and 481 developers respectively to examine usage patterns, avoidance reasons, and desired improvements.
Wang et al.~\cite{wang2023practitioners} combine interviews (15 professionals) and a survey (599 developers) to assess expectations around code completion.
McNutt et al.~\cite{mcnutt2023design} interview 15 data scientists on AI assistant design for notebooks, while Ziegler et al.~\cite{ziegler2022productivity} link Copilot usage metrics to self-reported productivity.

User studies provide insights into interactions.
Vaithilingam et al.~\cite{vaithilingam2022expectation} study 24 participants' usage of GitHub Copilot.
Barke et al.~\cite{barke2023grounded} observe 20 developers and identify ``acceleration'' and ``exploration'' modes.
Mozannar et al.~\cite{mozannar2024reading} track 21 participants and identify 12 Copilot-related activities.
Shah et al.~\cite{shah2025students} study student use of Copilot's chat in large codebases and propose the ``one-shot prompting'' interaction pattern.

Our study analyzes user reviews from the VS Code Marketplace to construct a comprehensive taxonomy of topics users discuss. 
Compared to the existing taxonomy by Zhou et al.~\cite{zhou2025exploring}, which focuses largely on dependability issues, our taxonomy not only encompasses their identified concerns but also offers broader coverage. 
Specifically, we uncover user-valued aspects and problems that their taxonomy overlooks. 
Notably, three of our top-level categories—general experience, pricing, and comparison—capture user sentiments that are absent from their framework. 
In addition, our findings verify certain contradictions in the current literature and challenge some existing conclusions, as discussed in Section~\ref{sec:results_rq2}.

% Unlike existing taxonomies that focus mainly on the problems developers encounter~\cite{zhou2025exploring}, ours offers broader coverage and scales to diverse content.

% Many study has evaluate the AI Coding Assistant~\cite{nguyen2022empirical}, such as GitHub Copilot.
% Nguyen et al.~\cite{nguyen2022empirical} evaluate correctness and understandability of Copilot's suggested code on 33 LeetCode questions and find .
% Yang et al.~\cite{yang2024robustness} interview 15 developers and 

% GitHub, Survey reveals AI’s impact on the developer experience, 2024, https://github.blog/2023-06-13-survey-reveals-ais-impact-on-the-developerexperience/. (Accessed November 2024).

\subsection{Mining App Reviews}
\label{subsec:app_review_analysis}

App stores have become vital platforms for software distribution, shaping engineering practices by connecting developers and users~\cite{dkabrowski2022analysing, al2019app}. While prior research has extensively analyzed app reviews through information extraction~\cite{vu2015mining}, classification~\cite{chen2014ar}, sentiment analysis~\cite{guzman2014users}, and clustering~\cite{pagano2013user}.
Classification and clustering are particularly relevant to our study. Classification assigns reviews to predefined categories—such as user intentions~\cite{maalej2016automatic} or functional versus non-functional requirements~\cite{jha2019mining}—while clustering uncovers themes or shared app characteristics without predefined labels~\cite{pagano2013user, guzman2014users}. Both manual~\cite{pagano2013user} and automated~\cite{chen2014ar} approaches have been widely adopted.

\section{Collecting AI Coding Assistants}
\label{sec:dataset}

% This section outlines the dataset collection process in Section~\ref{subsec:data_collection} and provides a brief analysis in Section~\ref{subsec:analysis_dataset}.

\subsection{VS Code AI Extension Collection}
\label{subsec:data_collection}

We begin by defining what constitutes an AI extension in the VS Code Extension Marketplace~\cite{vscode_marketplace}.
An extension is considered an AI extension if it includes at least one AI component that assists its functionality.
For instance, the GitHub Copilot Extension~\cite{github.copilot} is considered an AI extension, as it leverages the AI model to provide intelligent code completions or suggestions while developers code.
In contrast, we do not classify Azure Machine Learning as an AI extension, as it serves primarily as a tool for developing AI—supporting the building, training, and deployment of machine learning models—rather than incorporating AI into its functionality.

\begin{figure*}[]
  \centering
  \includegraphics[width=0.95\linewidth]{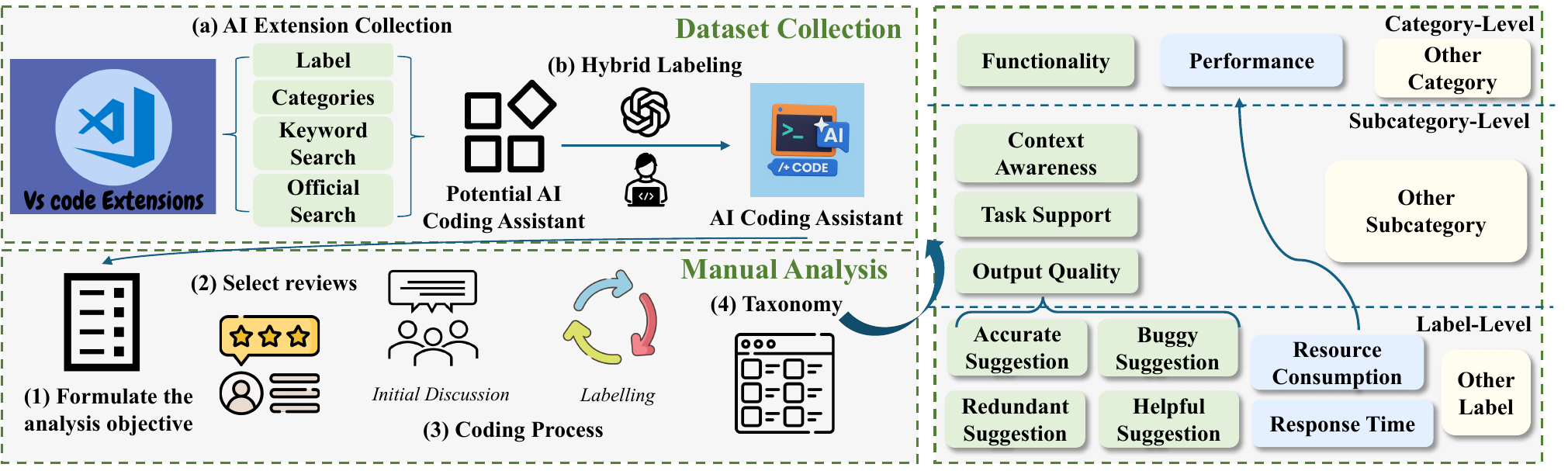}
  \caption{Our five-step process for analyzing reviews from the VS Code marketplace: (1) formulate analysis objective, (2) select reviews for analysis, (3) specify unit of analysis, (4) perform coding process, and (5) analyze dataset.}
  \label{fig:workflow}
\end{figure*}

As shown in the top-left of Fig.~\ref{fig:workflow}, we follow a two-step process to collect AI extensions from the VS Code Marketplace: (1) collecting AI extensions and (2) hybrid labeling.
We collect all potential AI extensions during the first step, and then we confirm the AI extensions in the second step.
% We collect all potential AI extensions use multiple methods: (1) Official search; (2) Categories; (3) Tags; (4) Description.
% Then we use a manual and automatic combined labeling process to label the AI extensions.
% Specifically, we first label a small number of extensions manually, and then find the best prompt for the GPT-4 model to label the rest of the extensions.

\textbf{AI Extension Collection.}
We collect metadata for 66,053 extensions available on the Microsoft VS Code Marketplace~\cite{vscode_marketplace} as of November 28, 2024.
To identify AI-related extensions, we examine several sources of information, including metadata (tags and categories) and extension descriptions.
For example, the GitHub Copilot Extension~\cite{github.copilot} includes tags such as ``ai'' and ``openai,'' and belongs to the AI category. 
Its description further states: \textit{``GitHub Copilot provides autocomplete-style suggestions from an AI pair programmer as you code,''} which strongly suggests the presence of AI capabilities.
In addition to these sources, the VS Code Marketplace provides an official search feature~\cite{vscode_marketplace_ai_search}, which returns extensions related to the keyword ``AI.''

To ensure comprehensive coverage, we adopt a multi-faceted approach to collect potential AI extensions using the following four methods:
(1) \textit{Official Search.}
We use the official Marketplace search feature with the keyword ``AI'', resulting in 1,623 candidate extensions.
(2) \textit{Category-Based Selection.}
Each extension in the Marketplace includes predefined categories in its metadata.
There are 20 categories in total. %, such as Other, Themes, Programming Languages, and Snippets.
Two authors review and select the categories `AI', `Chat', and `Machine Learning' as indicative of AI functionality. 
This approach yields 740 additional candidate extensions.
(3) \textit{Tag-Based Selection.}
Tags offer finer-grained metadata about extensions. 
We find a total of 49,733 unique tags across all extensions. 
Focusing on the tags that appeared more than 100 times, two authors select nine tags likely to indicate AI content: ai, gpt, chatgpt, openai, copilot, autocomplete, intellisense, deep learning, and machine learning. 
Extensions containing any of these tags contribute 1,052 entries to our candidate pool.
(4) \textit{Descriptions Keyword Matching.}
We download the descriptions of all extensions and apply keyword matching using a list of 55 AI-related terms, adapted and refined from prior work~\cite{hou2023large, liu2025protect}
The full list is available in the replication package.
Matching these keywords against extension descriptions results in 715 candidates.

After aggregating results from the four collection methods and removing duplicates, we identified a final set of 1,962 potential AI extensions. 
The individual contributions were as follows: 75 from the official search, 345 from categories, 559 from tags, and 414 from descriptions.
Searching only through the official search may not identify all AI extensions.

\textbf{Hybrid Labeling.}
We first randomly sampled 322 extensions from the pool of candidate AI coding assistants.
The sample size is calculated using a sample-size calculator~\cite{SurveyMonkey_SampleSize}, based on a 95\% confidence level and a 5\% margin of error, a setting commonly adopted in previous work~\cite{hata20199, lyu2024evaluating, wang2021restoring}.
Two authors independently label the sampled extensions as either confirmed AI extensions or not, based on predefined criteria.
Each extension's description is reviewed, and documentation is consulted when necessary.
The labeling achieves a Cohen's Kappa coefficient of 0.92~\cite{mchugh2012interrater}, indicating a high level of agreement~\cite{landis1977measurement}. 
Discrepancies are resolved through discussion.
In total, among the 322 sampled extensions, 195 are confirmed as AI extensions, while 127 are not.

To scale the labeling process, we explore the capabilities of GPT-4o. 
We input the extension name and description into GPT-4o and ask it to determine whether the extension qualifies as an AI extension.
We experiment with various prompting strategies.
% , including zero-shot and few-shot learning, as well as different prompt templates.
On the manually labeled set, we find that a zero-shot learning approach provides good performance, achieving a precision of 96.37\% and recall of 96.88\%.
Manual inspection of misclassifications shows that most errors stem from vague or insufficient descriptions; none involve highly installed or highly rated extensions.
We put the prompt used in this study in the replication package due to the limited space.
Using the final prompt and setting the temperature to 0 for deterministic output, we label the remaining potential extensions with GPT-4o.
This results in 1,085 confirmed AI extensions—55.3\% of the initial candidate set—representing 1.64\% of all extensions on the VS Code Marketplace.

\subsection{Dataset Overview}
\label{subsec:analysis_dataset}

% The VS Code marketplace hosts 66,053 extensions, including 1,085 AI extensions (1.64\%). 
Despite the small share of AI extensions (only 1.64\% of all extensions), we notice that AI extensions receive more attention, with an average of 124K installs, double that of non-AI extensions (57K).
Additionally, AI extensions receive significantly more feedback, with an average of 7.48 ratings per extension compared to 1.76 for non-AI extensions (p < 0.05, Mann-Whitney U test~\cite{mcknight2010mann}).
Both AI and non-AI extensions show similar inequality in installs and ratings, as indicated by high Gini coefficients~\cite{gini1912variability} (0.986 vs. 0.982 for installs; 0.931 vs. 0.873 for ratings). 
The Gini coefficient quantifies statistical dispersion, with 0 indicating perfect equality and 1 representing maximal inequality.
However, attention with AI extensions is even more concentrated: the top 10 most-installed AI extensions account for 86\% of total installs, and the top 30 most-rated receive 75\% of all user ratings.
This indicates that a small number of AI extensions dominate the market.

% As shown in Fig~\ref{fig:percentage_distribution}, 
AI extensions have seen rapid growth in recent years, with 90.2\% released between 2022 and 2024.
Nearly half appeared in 2023 (44.6\%) and 2024 (45.6\%), highlighting their recent surge.
In contrast, non-AI extensions have followed a more gradual distribution, with moderate growth in recent years (17.8\% in 2023, 18.2\% in 2024) and a significant portion released in earlier years.
This contrast underscores the sudden rise of AI-powered extensions, whereas non-AI extensions have evolved at a steadier pace over time.

\section{Study Design}
\label{sec:study_design}

% In this section, we first describe the research questions and then discuss the study design to answer these questions.

  % \item \textit{RQ1. What are the functionality of AI extensions?}
  % \item \textit{RQ3. What do users discuss in the Q \& A and reviews of AI extensions?}
  % \item \textit{RQ4. What factors contribute to the popularity of AI extensions in the VS Code marketplace?}
  % \item \textit{RQ5. What is the lifecycle of AI extensions in the VS Code marketplace?}
  % \item \textit{RQ6. What is the deprecation of AI extensions in the VS Code marketplace?}
% In first research question, we want to know what are these AI extensions.
% Specifically, we want to know what's there functionalities, 
% In RQ1, specifically , We want to know  in terms of popularity, maintenance, and security?
% In RQ3, specifically we want to answer three questions: (1) What do users discuss in the Q \& A? (2) What do users discuss in the reviews? (3) What are the differences between Q \& A and reviews?

% While each identified topic could warrant a dedicated study and deeper analysis, here we provide initial answers for each and explore opportunities for future research.

\subsection{Research Questions}
\label{subsec:research_questions}

% including the topics they discuss and the aspects they care about, appreciate, or dislike.
% \footnote{In this paper, the terms AI coding assistants, AI extensions, and assistants are used interchangeably to refer to AI-powered extensions in the IDE, such as GitHub Copilot.}

We aim to understand user perceptions of AI coding assistants by address the following research questions:

\begin{itemize}[leftmargin=10pt]
  \item \textbf{RQ1. What do users discuss about AI coding assistants?}
  \item \textbf{RQ2. What aspects of AI coding assistants do users care about, appreciate, and dislike?}
\end{itemize}

RQ1 seeks to identify and categorize the main topics users discuss about AI coding assistants.
RQ2 focuses on understanding user preferences, including what aspects of AI coding assistants they care about, appreciate, or dislike.
To build the taxonomy and systematically understand user preferences, we follow the process suggested by D\k{a}browski et al.~\cite{dkabrowski2022analysing} for analyzing app reviews in software engineering.
The process consists of the following steps, as illustrated in the lower-left of Fig.\ref{fig:workflow}: (1) formulate the analysis objective; (2) select reviews for analysis; (3) specify the unit of analysis; (4) perform the coding process; and (5) analyze the dataset.

\subsection{Research Object}
\label{subsec:research_object}

\textit{Formulate the analysis objective.}
To ensure the extraction of meaningful insights, we focus on a strategically selected subset of AI extensions rather than the full set collected in Section~\ref{sec:dataset}.
Specifically, we target two types of AI extensions with broad user reach and engagement: (1) those with the highest number of installs, and (2) those with the highest number of user reviews.
For the first criterion, we select the 30 most-installed AI extensions, which collectively account for 95.3\% of total installs and 68.5\% of all user reviews.
For the second, we include all AI extensions with more than 100 user reviews, as they demonstrate high user engagement. 
This adds two additional extensions not included in the top 30 by install count.
In total, we analyze 32 AI extensions, balancing popularity (install count) and user engagement (number of reviews) to support a comprehensive qualitative study.

% During preliminary review, we observed that extensions with low engagement—measured by the number of installs and user reviews—often contain vague or uninformative feedback (e.g., \textit{``superb extension''}), which offers limited analytical value.
% Following a similar approach to prior work~\cite{nahar2025product}, we select 32 AI extensions for qualitative analysis. 
% This set includes the 30 most-installed AI extensions, which collectively account for 95.3\% of total installs and 68.5\% of all user reviews. 
% These extensions are therefore highly representative of widely adopted AI extensions in the VS Code marketplace.
% In addition, we identify all AI extensions with more than 100 user reviews. 
% Two of these are not among the top 30 by install count but exhibit high user engagement. 
% To ensure comprehensive coverage of popular and actively discussed extensions, we include these two extensions as well.
% In total, our qualitative analysis is based on user reviews from 32 AI extensions.

\textit{Select reviews for analysis.}
We analyze user feedback to identify the topics users discuss regarding AI coding assistants. 
From a total of 5,908 reviews across 32 AI extensions, we exclude 649 reviews that contain fewer than 10 words, as such short reviews are unlikely to provide meaningful insights.
This results in 5,259 reviews.
Given the large volume, analyzing all 5,259 reviews is impractical. 
Therefore, we randomly sample 361 reviews, ensuring a 95\% confidence level with a 5\% margin of error.
This approach is consistent with prior studies~\cite{wang2021restoring, hata20199}.
Feedback in all languages is included to capture diverse perspectives (as we discussed in Section~\ref{sec:results_rq2}, users provide region- or country-specific comments), and translate non-English reviews into English using Google Translate.

\textit{Specify unit of analysis.}
We use the individual review as the unit of analysis.
Each review serves as a standalone piece of feedback that conveys a complete user experience, allowing for consistent manual interpretation and coding.

\subsection{Coding Process}
\label{subsec:coding_process}

We structure the coding process in two stages: an \textit{initial discussion} to establish coding principles, followed by \textit{iterative coding} to develop and refine the taxonomy.

\textit{Initial discussion.}
Before initiating substantive coding, four participants-three PhD candidates and one senior researcher with expertise in SE and AI-jointly reviewed a sample of 30 user reviews drawn from the full set of 361.
Through discussion, we establish four guiding principles:
(1) \textit{Multi-label annotation:}
A single review often addresses multiple independent themes—for instance, praising an assistant's helpful suggestions while criticizing its resource consumption. 
Therefore, we allow multiple labels per review when separate topics are discussed.
(2) \textit{Three-level taxonomy:}
A two-level hierarchy (category–leaf node), as used in prior work~\cite{zhou2025exploring}, proves insufficient for capturing the content observed in our pilot sample—too many leaf nodes fall under a single category, resulting in overly broad themes. 
We therefore adopt a three-level taxonomy (category–subcategory–leaf node) to better represent topic granularity, in line with prior research~\cite{humbatova2020taxonomy}.

(3) \textit{Bottom-up merging process:}
We adopt a bottom-up method to construct the taxonomy similar to prior work~\cite{humbatova2020taxonomy}.
In each round, two annotators independently assign one or more labels to each review. 
They then meet to reconcile any disagreements and consolidate semantically similar labels into unified expressions, which we refer to as leaf nodes. 
These leaf nodes are initially grouped directly under top-level categories, forming a category–leaf node structure. 
A subcategory layer is introduced only when a category contains a sufficiently diverse set of leaf nodes—typically more than five—resulting in a three-level structure (category–subcategory–leaf node).

(4) \textit{Hybrid card sorting:}
We adopt a hybrid card-sorting approach to provide an initial structure while retaining flexibility for refinement.
Specifically, we predefine five top-level categories—\textit{Functionality}, \textit{Usability}, \textit{Dependability}, \textit{Supportability}, and \textit{Performance}—based on established taxonomies from the app review literature~\cite{jha2019mining, kurtanovic2017mining}.
This choice is motivated by two considerations. 
First, app review mining is a mature field, and its taxonomies are derived from large-scale datasets covering diverse software types, offering generalizable top-level categories. 
Second, when reviewing our pilot sample, we found that most user comments naturally fit within these predefined dimensions. Without such a structure, our initial coding produced 11 top-level categories from just 30 reviews, resulting in a fragmented and inconsistent taxonomy.
All remaining subcategories and leaf nodes are derived through bottom-up consolidation, with annotators free to revise or expand the taxonomy as needed throughout the coding process.
%  those developed for large-scale ecosystems such as the Apple App Store.

% We adopt a hybrid card-sorting approach rather than a fully open-ended method, as it will lead to a more organized taxonomy.
% Specifically, we predefine five top-level Categories—Functionality, Usability, Dependability, Supportability, and Performance—drawing on well-established taxonomies from the app review literature, particularly those developed for large-scale ecosystems such as the Apple App Store.
% These predefined dimensions provide a high-level structural scaffold.
% In addition, we include one initial Subcategory—Suggestion Content—informed by Zhou et al.'s~\cite{zhou2025exploring} analysis of user feedback on GitHub Copilot. 
% Our pilot review of the initial 30 samples confirms that this seed taxonomy covers most observed topics. 
% Beyond this, all additional Subcategories and Leaf nodes emerge through bottom-up consolidation. 
% Annotators retain full flexibility to revise or rename all levels of the taxonomy as needed throughout the process.

\textit{Iterative coding.}
We conduct annotation in iterative batches of 50 user reviews. 
In each round, two annotators independently label reviews based on principles set in the initial discussion.
After three rounds, we identify a gap in the taxonomy: while technical and functional aspects are well covered, broader sentiments—such as perceptions of productivity or satisfaction—are underrepresented.
To address this, we revisit earlier reviews to ensure no content related to users' general impressions is overlooked, resulting in three new categories: \textit{general experience} (e.g., productivity boost), \textit{pricing} (e.g., free to use), and \textit{comparison} (e.g., evaluation against other tools).
By the end of Round 3, the taxonomy includes 8 categories, 16 subcategories, and 55 leaf nodes.

In Rounds 4 through 6, the taxonomy stabilizes. 
Only five additional leaf nodes emerge across the three rounds (1, 2, and 2, respectively), and no new categories or subcategories are introduced—indicating structural saturation~\cite{humbatova2020taxonomy}. 
At this stage, we finalize the taxonomy and compile a formal annotation guideline, including definitions and illustrative examples for each category, subcategory, and leaf node.

The two annotators independently label the remaining 17\% of the dataset. 
Disagreements are resolved through discussion. The inter-rater agreement for this final phase is K = 0.983, indicating strong consistency between annotators~\cite{landis1977measurement}. 
Only two additional labels are introduced in this phase.
In total, we annotated 361 user reviews, resulting in 62 unique leaf nodes, structured across 16 subcategories and 8 top-level categories.
Note: In the later content, ``label’’ refers to the leaf nodes.

\subsection{Sentiment Analysis}
\label{subsec:sentiment_analysis}

A single review may express multiple sentiments—for example, a user may praise helpful suggestions but criticize high memory consumption.
Therefore, instead of assigning one sentiment per review, we analyze sentiment at the leaf node level.
Although leaf nodes are finalized through bottom-up consolidation, sentiment is annotated when individual labels are first applied.
Each label is marked as like, dislike, or neutral (including feature requests or emotionally neutral mentions).
In the first six rounds, two annotators jointly labeled sentiment during coding.
In the final phase (17\% of reviews), sentiment was annotated independently.
We report inter-rater agreement only for this phase, as earlier rounds involved evolving labels and collaborative decisions.
Agreement for the final phase is K = 0.943, indicating high consistency~\cite{landis1977measurement}.

\section{Categorize discussion of coding assistants}
\label{sec:results_rq2}

In this section, we present the taxonomy derived from our analysis of user reviews of AI coding assistants in VS Code marketplace. 
The taxonomy results are presented in Table~\ref{tab:user_feedback_categories}.
We categorize topics of user reviews into eight high-level categories: \textit{functionality}, \textit{usability,} \textit{dependability}, \textit{system performance}, \textit{supportability}, \textit{general experience}, \textit{pricing}, and \textit{comparison}.
In total, we identify 8 categories, 16 subcategories, and 62 labels.
In the remainder of this subsection, we define each category along with its corresponding subcategories and labels.
% A more detailed discussion of what users care about, as well as what they like or dislike, is provided in Section~\ref{subsec:results_like_dislike}.
We report the frequencies leaf nodes using the multiplication symbol ($\times$).

% 定义情感条形图宏
\newcommand{\sentimentbar}[3]{%
  \begin{tikzpicture}[baseline={(0,-0.09)}]  % 修改基线设置
    % 明确定义边界框以控制总宽度
    \path[use as bounding box] (-0.9,-0.15) rectangle (2.7,0.15);
    
    % 创建基线参考点
    \coordinate (baseline) at (0,0);
    
    % 左侧百分比 - 调整垂直位置
    \node[anchor=east, font=\scriptsize, text width=0.7cm, align=right] at (-0.1,0) {#1\%};
    
    % 条形图 - 使用正确的十六进制颜色语法
    \definecolor{likecolor}{HTML}{42b481}
    \definecolor{requestcolor}{HTML}{dcdcdc}
    \definecolor{dislikecolor}{HTML}{ec5b60}
    
    \fill[likecolor] (0,-0.1) rectangle (#1/50,0.1);
    \fill[requestcolor] (#1/50,-0.1) rectangle ({(#1+#3)/50},0.1);
    \fill[dislikecolor] ({(#1+#3)/50},-0.1) rectangle (2,0.1);
    
    % 右侧百分比
    \node[anchor=west, font=\scriptsize, text width=0.7cm, align=left] at (2.1,0) {#2\%};
  \end{tikzpicture}%
}

\newcommand{\fulllikebar}{%
  \begin{tikzpicture}[baseline={(0,-0.09)}]  % 修改基线设置
    % 明确定义边界框以控制总宽度
    \path[use as bounding box] (-0.9,-0.15) rectangle (2.7,0.15);
    % 创建基线参考点
    \coordinate (baseline) at (0,0);
    % 左侧百分比
    \node[anchor=east, font=\scriptsize, text width=0.7cm, align=right] at (-0.1,0) {100\%};
    
    % 定义颜色
    \definecolor{likecolor}{HTML}{42b481}
    
    % 条形图
    \fill[likecolor] (0,-0.1) rectangle (2,0.1);
    % 右侧百分比
    \node[anchor=west, font=\scriptsize, text width=0.7cm, align=left] at (2.1,0) {0\%};
  \end{tikzpicture}%
}

\newcommand{\fulldislikebar}{%
  \begin{tikzpicture}[baseline={(0,-0.09)}]  % 修改基线设置
    % 明确定义边界框以控制总宽度
    \path[use as bounding box] (-0.9,-0.15) rectangle (2.7,0.15);
    % 创建基线参考点
    \coordinate (baseline) at (0,0);
    % 左侧百分比
    \node[anchor=east, font=\scriptsize, text width=0.7cm, align=right] at (-0.1,0) {0\%};
    
    % 定义颜色
    \definecolor{dislikecolor}{HTML}{ec5b60}
    
    % 条形图
    \fill[dislikecolor] (0,-0.1) rectangle (2,0.1);
    % 右侧百分比
    \node[anchor=west, font=\scriptsize, text width=0.7cm, align=left] at (2.1,0) {100\%};
  \end{tikzpicture}%
}

\begin{table*}[t]
  % \small
  \caption{
    Categorization of discussion of AI Coding Assistants. 
    PL, DEI are abbreviations for Programming Language and Development Environment Integration.
    The table is sorted by the number of reviews in descending order.
    The last column represents users’ sentiment toward each category, subcategory, or label.
  }
  \centering
  \setlength{\tabcolsep}{3pt}  % 减少列间距以节省空间
  \resizebox{0.85\textwidth}{!}{
  \begin{tabular}{cl|l|r|r|>{\centering\arraybackslash}p{3.6cm}}
    \toprule
    \textbf{ID} & \textbf{Category-Subcategory} & \textbf{Description} & \textbf{No.} & \textbf{Rate} & \textbf{Sentiment} \\
    \midrule
    \multirow{7}{*}{1} & \textbf{Functionality} 
    & Core code-generation and assistance features. & 238 & 32.2\% 
    & \sentimentbar{64}{31}{5} \\
    & \quad \textbullet{} Suggestion Content
    & Opinions on code suggestion content. & 101 & 13.7\% 
    & \sentimentbar{59}{37}{4} \\
    & \quad \textbullet{} PL, Library, Task Support. 
    & Support for languages, libraries, and SE tasks. & 73 & 9.9\% 
    & \sentimentbar{78}{17}{5} \\
    & \quad \textbullet{} Understanding ability 
    & Ability to understand code and user intent. & 45 & 6.1\% 
    & \sentimentbar{62}{29}{9} \\
    & \quad \textbullet{} Context Awareness 
    & Ability to leverage code/project context. & 21 & 2.8\% 
    & \sentimentbar{38}{57}{5} \\
    & \quad \textbullet{} DEI 
    & Integration with tools (terminal, debugger). & 7 & 1.0\% 
    & \fulllikebar \\
    \midrule
    \multirow{5}{*}{2} & \textbf{General Experience} 
    & Overall experience and emotional response. & 136 & 18.4\% 
    & \sentimentbar{90}{10}{0} \\
    & \quad \textbullet{} Productivity 
    & Reported acceleration or slow down of coding speed, flow. & 83 & 11.2\%  
    & \sentimentbar{90}{10}{0} \\
    & \quad \textbullet{} General Discussion 
    & Open-ended reflections or pure praise and claims. & 28 & 3.8\% 
    & \sentimentbar{89}{11}{0} \\
    % & \quad \textbullet{} Pure & & 25 & 3.4\% 
    % & \sentimentbar{92}{8}{0} \\
    & \quad \textbullet{} Helpfulness 
    & Usefulness of suggestions solving problems. & 23 & 3.4\% 
    & \sentimentbar{92}{8}{0} \\
    % & \quad \textbullet{} Slow down & & 8 & 1.1\% 
    % & \fulldislikebar \\
    \midrule
    \multirow{4}{*}{3} & \textbf{Usability} 
    & Interface design and ease of use. & 104 & 14.1\%
    & \sentimentbar{53}{36}{12} \\
    & \quad \textbullet{} UI \& Interactivity 
    & Interface layout, chat panel, pop-up quality, cursor control. & 60 & 8.1\% 
    & \sentimentbar{55}{30}{15} \\
        & \quad \textbullet{} Controllability 
    & Settings, model choice, interruption control. & 21 & 2.8\% 
    & \sentimentbar{76}{19}{5} \\
    & \quad \textbullet{} Learnability 
    & On-boarding difficulty and docs clarity. & 16 & 2.2\% 
    & \sentimentbar{38}{50}{13} \\
    & \quad \textbullet{} Predictability 
    & Consistency of responses, avoids corruption. & 7 & 1.0\% 
    & \fulldislikebar \\
    \midrule
    \multirow{6}{*}{4} & \textbf{Dependability} 
    & Trustworthiness: reliability, security, ethics, uptime. & 83 & 11.2\% 
    & \sentimentbar{19}{77}{4} \\
    & \quad \textbullet{} Reliability 
    & Stability, crashes, install failures, fallbacks. & 37 & 5.0\% 
    & \sentimentbar{13}{84}{3} \\
    & \quad \textbullet{} Legal \& Ethical Concerns 
    & License compliance, AI-ethics concerns. & 23 & 3.1\% 
    & \sentimentbar{22}{74}{4} \\
    & \quad \textbullet{} Security \& Privacy 
    & Risk of code leaks, privacy-safeguard adequacy. & 12 & 1.6\% 
    & \sentimentbar{17}{83}{0} \\
    & \quad \textbullet{} Availability 
    & Offline capability and regional limitations. & 10 & 1.4\% 
    & \sentimentbar{40}{50}{10} \\
    \midrule
    \multirow{3}{*}{5} & \textbf{Pricing} 
    & Monetary cost, free tiers, perceived value. & 55 & 7.4\% 
    & \sentimentbar{69}{29}{2} \\
    & \quad \textbullet{} Free to use 
    & Positive reactions to generous free tiers. & 32 & 4.3\% 
    & \sentimentbar{97}{0}{3} \\
    & \quad \textbullet{} Value Perception 
    & Overpricing claims versus fair-price praise. & 23 & 3.1\%
    & \sentimentbar{68}{32}{0} \\
    % & \quad \textbullet{} Overpriced for features & & 15 & 2.1\% 
    % & \fulldislikebar \\
    % & \quad \textbullet{} Money worth for the service & & 7 & 1.0\% 
    % & \fulllikebar \\
    % & \quad \textbullet{} Remove Education Plan & & 1 & 0.1\% 
    % & \fulldislikebar \\
    \midrule
    \multirow{5}{*}{6} & \textbf{Supportability} 
    & Post-deployment support, compatibility, rollout. & 48 & 6.5\% 
    & \sentimentbar{58}{33}{8} \\
    & \quad \textbullet{} Compatibility 
    & OS / IDE / web compatibility, conflicts. & 18 & 2.4\% 
    & \sentimentbar{47}{35}{18} \\
    & \quad \textbullet{} Serviceability 
    & Vendor or community support responsiveness. & 15 & 2.0\% 
    & \sentimentbar{93}{7}{0} \\
    & \quad \textbullet{} Feature Availability 
    & Speed and openness of new-feature access. & 11 & 1.5\% 
    & \sentimentbar{46}{46}{9} \\
    & \quad \textbullet{} Maintainability 
    & Stability across updates, maintenance pace. & 4 & 0.5\% 
    & \sentimentbar{20}{80}{0} \\
    \midrule
    \multirow{4}{*}{7} & \textbf{Comparison} 
    & Comparisons with other AI tools. & 44 & 6.0\% 
    & \sentimentbar{71}{27}{2} \\
    & \quad \textbullet{} With Competition 
    & General comparisons to rival products. & 21 & 2.9\% 
    & \sentimentbar{81}{19}{0} \\
    & \quad \textbullet{} With Github Copilot 
    & Contrasts in accuracy, speed, price vs Copilot. & 19 & 2.6\% 
    & \sentimentbar{74}{21}{5} \\
    & \quad \textbullet{} With GPT 
    & Evaluations versus raw ChatGPT / GPT-4. & 4 & 0.5\% 
    & \fulldislikebar \\
    \midrule
    \multirow{5}{*}{8} & \textbf{Performance} 
    & Latency, throughput, resource footprint issues. & 31 & 4.2\% 
    & \sentimentbar{58}{42}{0} \\
    & \quad \textbullet{} Response Time 
    & Perceived waiting time between request and output. & 17 & 2.3\% 
    & \sentimentbar{82}{18}{0} \\
    & \quad \textbullet{} Resource Consumption 
    & CPU, memory, battery drain complaints. & 9 & 1.2\% 
    & \sentimentbar{22}{78}{0} \\
    & \quad \textbullet{} Rate Limiting 
    & Complaints about request caps & 5 & 0.7\% 
    & \sentimentbar{40}{60}{0} \\
    % & \quad \textbullet{} Throughput & & 1 & 0.1\% 
    % & \fulldislikebar \\
    \midrule
    \multicolumn{6}{c}{
      \begin{tikzpicture}[baseline={(0,0)}]
        \definecolor{likecolor}{HTML}{42b481}
        \definecolor{neutralcolor}{HTML}{dcdcdc}
        \definecolor{dislikecolor}{HTML}{ec5b60}
        
        \fill[likecolor] (0,0) rectangle (0.5,0.25);
        \node[right] at (0.55,0.125) {Like};
        
        \fill[neutralcolor] (2,0) rectangle (2.5,0.25);
        \node[right] at (2.55,0.125) {Request};
        
        \fill[dislikecolor] (4,0) rectangle (4.5,0.25);
        \node[right] at (4.55,0.125) {Dislike};
      \end{tikzpicture}
    } \\
    \bottomrule
  \end{tabular}}
  \label{tab:user_feedback_categories}
\end{table*}

% The categories described above focus on the specific features of AI coding assistants, while the following categories reflect users' overall perceptions and subjective experiences.
% \textit{General experience} captures users' overarching impressions of the AI coding assistant, including its perceived helpfulness, impact on productivity, and general remarks such as praise or criticism.
% \textit{Pricing} includes feedback related to cost, such as whether the tool is free to use and users' perceptions of its value.
% \textit{Comparison} refers to reviews in which users compare the AI coding assistant with alternative tools, such as GitHub Copilot, ChatGPT, or other competing extensions.

\subsection{Functionality}
\label{subsec:functionality}

The functionality category refers to user discussions about the core features of the AI coding assistant, including five subcategories: Suggestion Content, Language, Library, Task Support, Context Awareness, and Understandability, and Development Environment Integration.
It has a total of 18 labels.

% \vspace*{0.2cm}
% \noindent 
\subsubsection{Suggestion Content}
% \textbf{\textit{Suggestion Content.}} 
This subcategory covers a set of users' opinions on suggestions from the coding assistants.
The most frequently discussed label is \textit{accuracy of suggestions (43$\times$)}, reflecting how suggestions match the user's expectations. 
The second most discussed label is \textit{helpfulness of suggestion (32$\times$)}, indicating whether the suggestion is useful and aligns with the user's intent—even if not entirely accurate, it may still provide valuable hints.
The third most mentioned label is \textit{completeness and redundancy of suggestions (11$\times$)}, which concerns whether the output is sufficiently complete or contains unnecessary parts. 
This is followed by \textit{failure on complex tasks (5$\times$)}, referring to the assistant's limitations when handling more sophisticated programming problems.
Next is \textit{buggy suggestion (4$\times$)}, which captures instances where the AI produces incorrect or malfunctioning code. 
Finally, \textit{other (3$\times$)} includes less common issues such as hallucinated content or severely outdated suggestion.

% \vspace*{0.2cm}
% \noindent 
\subsubsection{PL, Library, Task Support}
% \textbf{\textit{PL, Library, Task Support.}} 
This subcategory reflects user discussions about the AI coding assistant's ability to support specific programming languages (PL), libraries, and tasks.
The most discussed label is \textit{task support (36$\times$)}, which focuses on capabilities beyond basic code completion—such as test generation, code refactoring, and code summarization. 
We exclude general code generation since it is a standard feature across all assistants analyzed.
The second most discussed label is \textit{PL support (28$\times$)}, with users referencing 13 languages.
The third discussed label is \textit{library or framework support (9$\times$)}.

% The most discussed label is \textit{Task support (36$\times$)}.
% Here we only the label besides code generation as all the AI coding assistants we analysis have the basic code generation function.
% We want to understand what's users' need beyond just code generation.
% Users discuss about the test generation, code refactoring, code summarization etc. 
% The second most discussed label is \textit{PL support 28$\times$}.
% Here 13 PL are mentioned by users, including Java, Python, C++, JavaScript, etc.
% The third most discussed label is \textit{Library or framework support (9$\times$)}.

\subsubsection{Understandability}
% \textbf{\textit{Understandability.}}
his subcategory captures users’ concerns about the assistant’s ability to comprehend code, errors, and human intent after retrieving the relevant context. 
Users frequently highlight the importance of  \textit{contextual understanding (23$\times$)}, \textit{code understanding (8$\times$)}, \textit{human instruction understanding (6$\times$)}, \textit{good explanation of the code (4$\times$)}, and \textit{error or syntax understanding (3$\times$)}, referring to the assistant's ability to identify issues in the code.

% \vspace*{0.2cm}
% \noindent 
\subsubsection{Context Awareness}
% \textbf{\textit{Context Awareness.}}
his subcategory reflects users’ expectations regarding the assistant’s ability to accurately retrieve and maintain contextual information.
\textit{Project/Codebase context awareness (13$\times$)} captures expectations for understanding and navigating across project- or repository-level contexts.
\textit{Context memory capacity (7$\times$)} refers to the ability to retain and leverage past interactions.

% \vspace*{0.2cm}
% \noindent 

% \vspace*{0.2cm}
% \noindent 
\subsubsection{Development Environment Integration}
% \textbf{\textit{Development environment integration.}}
Users mention it which refers to the assistant's ability to work seamlessly with parts of the development environment, including the terminal, version control systems, and debugging tools.

\subsection{Usability}
\label{subsec:usability}

This category includes reviews that discuss the user interface and ease of use of the AI coding assistant.
It includes four subcategories: UI \& interactivity, learnability, controllability, and predictability, with a total of 12 labels.

% \vspace*{0.2cm}
% \noindent 
% \textbf{\textit{UI \& Interactivity.}} 
\subsubsection{UI \& Interactivity}
This subcategory includes user feedback on interface design and interaction mechanisms.
The most discussed label is \textit{chat interface (23$\times$)}, where users comment on chat-based interaction, which differs from traditional code completion. 
For example, some users find in-line chat convenient.
\textit{General design and user experience (14$\times$)} captures feedback on intuitive, user-friendly interfaces and seamless access to AI functionality. 
\textit{Suggestion interface \& interaction (13$\times$)} includes feedback on how suggestions are presented (e.g., with stars, syntax highlighting, or snooze buttons) and how users interact with them (e.g., using shortcuts or whether suggestions are proactive).
\textit{Cursor interactions (6$\times$)} refers to feedback on cursor control, such as assistants changing cursor position unexpectedly or correctly predicting where the cursor should move.
\textit{Layout \& navigation (4$\times$)} includes comments about layout features like split view, sidebar management, and navigating to functions or class definitions.

% \vspace*{0.2cm}
% \noindent 
% \textbf{\textit{Learnability.}} 
\subsubsection{Learnability}
This subcategory reflects user discussions about whether the assistant is easy to learn and use.
\textit{On-boarding difficulty (6$\times$)} is the most discussed label, referring to challenges in getting started, including the learning curve and setup process.
\textit{Ease of use (5$\times$)} reflects user impressions that the assistant is intuitive and includes helpful features such as help bar.
\textit{Documentation (5$\times$)} includes comments on unclear or insufficient documentation.

% \vspace*{0.2cm}
% \noindent 
% \textbf{\textit{Controllability.}} 
\subsubsection{Controllability}
This subcategory reflects user feedback on how much control they have over the behavior and configuration of the AI coding assistant.
The most discussed label is \textit{Customization (12$\times$)}, where users discuss support for pre-configured and custom commands, configurability settings, and the ability to manage code snippets.
\textit{Model choice (4$\times$)} refers to the ability to select different models, including local models or cloud-based ones like GPT-4o.
\textit{Extension dependency issue (4$\times$)} includes comments on extensions being automatically installed due to dependencies, sometimes without user consent.

% \vspace*{0.2cm}
% \noindent 
% \textbf{\textit{Predictability.}} 
\subsubsection{Predictability}
Some users report that the assistant may \textit{mess up their existing code (6$\times$)}, introducing unexpected changes. Others mention \textit{inconsistent language output}, such as responses switching between English and French.

\subsection{Dependability}
\label{subsec:dependability}

This category includes user reviews that raise concerns about the dependability or trustworthiness of the assistant, covering aspects such as reliability, availability, and security.
It contains four subcategories: reliability, legal \& ethical concerns, security \& privacy, and availability.
% , with a total of twelve labels.

% \vspace*{0.2cm}
% \noindent 
% \textbf{\textit{Reliability.}} 
\subsubsection{Reliability}
This subcategory captures user feedback about the stability and consistent operation of the extension.
The most discussed label is \textit{bug of the extension (11$\times$)}, where users report crashes, unresponsiveness, or specific bugs.
\textit{Fallback to weak model (6$\times$)} includes complaints that the assistant downgrades to a weaker model or becomes less responsive over time, leading to a noticeable drop in answer quality.
\textit{Login issue (6$\times$)} refers to problems with logging in or authentication failures.
\textit{Installation issue (3$\times$)} includes cases where users are unable to install the extension or find it unusable after installation.
Another issue mentioned is \textit{AI service mistakenly blocking valid queries (2$\times$)}.

% \vspace*{0.2cm}
% \noindent 
% \textbf{\textit{Legal \& Ethical concerns.}}
\subsubsection{Legal \& Ethical concerns}
\textit{AI ethics (19$\times$)} is the most discussed label, where users raise concerns about open-source culture and AI ethics, such as being charged for models trained on their code.
\textit{License compliance (4$\times$)} refers to concerns generated code may violate code licenses.

% \vspace*{0.2cm}
% \noindent 
% \textbf{\textit{Security \& Privacy.}} 
\subsubsection{Security \& Privacy}
Users discuss \textit{privacy protection (6$\times$)}, such as the risk of data leakage, and \textit{permission over-reach (3$\times$)}, where extensions request excessive permissions.
One user also mentions concern about potential data being sent to external servers, raising security risks.

% \vspace*{0.2cm}
% \noindent 
% \textbf{\textit{Availability.}} 
\subsubsection{Availability}
Users express the need for consistent access to the assistant.
Some users complain or request that the extension should \textit{work locally (5$\times$)} without internet connection.
Another issue is \textit{region limitation (5$\times$)}, where users report that the extension is unavailable in their geographic area.

\subsection{Supportability}
\label{subsec:supportability}
This category includes user reviews concerning the assistant's ability to operate across different operating systems and IDEs (i.e., compatibility), as well as concerns related to updates and maintenance (e.g., maintainability).
It comprises three subcategories: compatibility, feature availability, and maintainability, with a total of nine labels.

% \vspace*{0.2cm}
% \noindent 
% \textbf{\textit{Compatibility.}} 
\subsubsection{Compatibility}
% This subcategory captures user discussions about the compatibility of the AI coding assistant.
\textit{IDE compatibility (17$\times$)} is the most discussed label, where users request support for additional IDEs, such as Atom.
\textit{Operating system compatibility (2$\times$)} includes comments about extending support to other platforms, such as Linux.
Some users also mention that the assistant should be \textit{compatible with the web version (2$\times$)} of the IDE.

% \vspace*{0.2cm}
% \noindent 
% \textbf{\textit{Feature Availability.}} 
\subsubsection{Feature Availability}
In this subcategory, users express appreciation for \textit{rapid feature improvements (11$\times$)} but also frustration with the \textit{waiting list for new features (5$\times$)}.

% \vspace*{0.2cm}
% \noindent 
% \textbf{\textit{Maintainability.}} 
\subsubsection{Maintainability}
Some users report concerns about the assistant's \textit{lack of maintenance (1$\times$)}, the \textit{frequency of updates (2$\times$)}, and instances where updates \textit{break functionality (2$\times$)}.

\textit{Serviceability (15$\times$)} is also frequently mentioned, referring to the development team's responsiveness to customer or community support requests.

% \vspace{-0.5em} % 调整为你希望的数值s
\subsection{System Performance}
\label{subsec:performance}
This category includes user reviews focused on the system performance of the AI coding assistant, such as response time, resource consumption, and rate limiting.
It contains three labels.
Users frequently discuss \textit{response time (17$\times$)}, expressing appreciation for fast performance while also complaining about occasional unresponsiveness.
\textit{Resource consumption (9$\times$)} is another common concern, with users reporting that the assistant consumes excessive system resources, such as high CPU and memory usage.
A few users also mention \textit{rate limiting (3$\times$)}, referring to restrictions on the number of requests or API calls the assistant can handle.

\subsection{Others}
\label{subsec:other}

\subsubsection{General Experience}
This category includes topics not directly related to the functional features of the AI coding assistant, but rather to users' overall experience.
It contains one subcategory (productivity) and five labels.

\textit{Productivity.}
As AI coding assistants are designed to enhance user productivity~\cite{vaithilingam2022expectation}, \textit{productivity (83$\times$)} is the most frequently discussed topic.
Users often report that the assistant helps \textit{boost their productivity (73$\times$)}, while some mention that it can \textit{slow down coding speed (8$\times$)}. 
A few users also state that it helps them \textit{find a solution to their problem (3$\times$)}.

In addition, some users engage in more \textit{general discussion (28$\times$)}, including reflections on mistake tolerance, concerns about over-reliance on the assistant, or pure \textit{praise or complaints (25$\times$)}.
Users also comment on the \textit{helpfulness (25$\times$)} of the assistant, noting that it supports developers at various stages of experience.

\subsubsection{Pricing}
% \label{subsec:pricing}
This category includes user reviews related to the pricing of the assistant.
Two labels are discussed.
Users often praise the assistant for being \textit{free to use (33$\times$)} and share opinions on its \textit{value perception (23$\times$)}, with some considering it overpriced and others finding it worth the cost.

\subsubsection{Comparison}
% \label{subsec:comparison}
This category includes user reviews that compare the AI coding assistant with other AI tools .
It contains three labels.
Users compare the assistant with general \textit{competitors (21$\times$)}, as well as specifically with \textit{GitHub Copilot (19$\times$)} and \textit{ChatGPT (4$\times$)}.

\section{What aspects of AI Coding Assistants Do Users Care About, Like, and Dislike?}
\label{subsec:results_like_dislike}

% In this section, we address the second research question. 
We summarize the insights from user reviews into six findings.
We quote original user review texts to support our findings.
Each review is referenced using an identifier (e.g., (R1, 5\ding{80}) refers to the first review in our dataset, where the user gave the AI extension a 5-star rating—the highest possible score in the VS Code Marketplace, which reflects the user’s evaluation of the extension itself (not the quality of the review), with 1 being the lowest and 5 the highest.

\begin{table}[]
  \setlength{\tabcolsep}{3pt}
  \caption{What Do Users Like and Dislike?}
  \centering
  \resizebox{0.9\linewidth}{!}{
  \begin{tabular}{llr|llr}
    \toprule
    \multicolumn{3}{c|}{\textbf{Top-15 Users' Like}} & \multicolumn{3}{c}{\textbf{Top-15 Users' Dislike}} \\
    \midrule
    no.\ 
      & \multicolumn{1}{c}{Label} 
      & N.
      & no.\ 
      & \multicolumn{1}{c}{Label} 
      & N. \\
    \midrule
    L1  & Accuracy suggestion            & 39  & 
    D1  & helpfulness suggestion       & 17 \\
    L2  & Task support                      & 24  & 
    D2  & AI ethics                        & 15 \\
    L3  & PL Support                        & 20  & 
    D3  & Bug of the extension             & 12 \\
    L4  & Chat interface                    & 16  & 
    D4  & complet \& redundant      & 8  \\
    L5  & helpfulness suggestion         & 14  & 
    D5  & Resource Consumption             & 7  \\
    L6  & Response Time                     & 14  & 
    D6  & Project Context Support          & 6  \\
    L7  & Serviceability                    & 14  & 
    D7  & Suggestion UI                    & 6  \\
    L8  & Customization                     & 12  & 
    D8  & Chat interface                   & 6  \\
    L9  & Code understanding                & 11  & 
    D9  & PL Support                       & 6  \\
    L10 & IDE Compatibility                 & 8   & 
    D10 & Context-memory capacity          & 6  \\
    L11 & General design                    & 7   & 
    D11 & On-boarding Difficulty           & 6  \\
    L12 & Project Context Support           & 7   & 
    D12 & Mess up the code                 & 5  \\
    L13 & Framework support                 & 6   & 
    D13 & Fallback to weak model           & 5  \\
    L14 & Suggestion UI                     & 6   & 
    D14 & Frustration waiting list    & 5  \\
    L15 & On-boarding Difficulty            & 5   & 
    D15 & Login Issue                      & 5  \\
    \bottomrule
  \end{tabular}}
  \label{tab:like_dislike}
  \vspace{-1.5em}
\end{table}

\vspace*{0.1cm}
\noindent
\textbf{Finding 1. Users generally perceive that AI coding assistants help boost productivity, though perceptions of helpfulness vary by experience level.}
Productivity has been a prominent topic since the introduction of GitHub Copilot~\cite{chen2021evaluating}.
Prior studies report perceived productivity gains among users~\cite{vaithilingam2022expectation, peng2023impact, martinovic2025perceived, imai2022github, wang2023practitioners}, though some lab-based experiments find no statistically significant improvements~\cite{mozannar2024reading,vaithilingam2022expectation}.
In our study, 90\% of reviews labeled with productivity express positive sentiment, describing how the assistant streamlines workflows, reduces repetitive tasks, and improves efficiency—for example: \textit{``not having to type every single repetitive function out or imports''} (R94, 5\ding{80}).
This aligns with the ``acceleration mode'' observed in prior work, where developers use the assistant to implement known solutions more quickly~\cite{barke2023grounded}.
Co-occurrence analysis between labels shows that productivity often appears alongside accurate suggestions (13$\times$), code understanding (7$\times$), PL support (6$\times$), and customization (6$\times$).
The first three are expected, as strong contextual understanding and language-specific support contribute to faster development.
Customization, on par with PL support, is somewhat unexpected. 
Upon review, we find that users value the ability to tailor the assistant to their needs, enhancing productivity:
\textit{``easily choose the languages you want to focus on this is so valuable''} or
\textit{``The ability to customize [assistant]'s suggestions and responses helps developers align the tool with their coding style and preferences...accelerating the coding process.''}
Indeed, customization is the 7\textsuperscript{th} most liked label, with 12 mentions (see Table~\ref{tab:like_dislike}).

Helpfulness has also been debated in prior research. 
Some argue that AI assistants benefit experienced developers who understand how to use them effectively~\cite{dakhel2023github}, while others suggest they are especially helpful to novices~\cite{prather2023s}.
Our findings support the latter: 14 out of 15 junior developers explicitly describe the assistant as helpful—e.g.,
\textit{``I am a beginner programmer and it is helping me a lot to build a project''} (R319, 5\ding{80}).
In contrast, experienced developers express more skepticism; two out of five report limited value:
\textit{``For anyone who really knows how to code, save yourself a lot of frustration''} (R14, 1\ding{80}).

\vspace*{0.1cm}
\noindent
\textbf{Finding 2. Users care most about suggestion content.
Accurate suggestions are highly valued, but users often criticize redundant, incomplete, or buggy outputs and express mixed views on the helpfulness of suggestions.}
As shown in Table~\ref{tab:user_feedback_categories}, suggestion content is the most frequently discussed topic, accounting for 14\% of all labeled mentions. 
Among these, 59\% of reviews express positive sentiment, while 41\% are negative.
A key aspect users appreciate is the \textit{accuracy of suggestions}, which received 39 positive mentions—making it the most liked label in our dataset (see Table~\ref{tab:like_dislike}). 
One user remarked:
\textit{``80\% less keyboard touching. Autocomplete is pure magic. Feels like it's connected directly to your mind''} (R164, 5\ding{80}).
However, users also frequently highlight frustrations, particularly regarding the \textit{helpfulness of suggestions} (17 dislikes), \textit{redundant or incomplete outputs} (8 dislikes), and \textit{buggy suggestions} (4 dislikes), including failures on complex tasks (4 dislikes). 
Notably, issues with redundancy and incompleteness have received little attention in prior studies, yet they represent the second most cited reason for negative sentiment within suggestion-related feedback. 
For example, users report: \textit{``Constantly barfs words on the screen 90+\% is repetitive``''} (R14, 1\ding{80}), and \textit{``it only predicts one character for me''} (R34, 1\ding{80}).

\vspace*{0.1cm}
\noindent
\textbf{Finding 3. Users concern about context awareness.  While AI assistants understand code well given context, they often struggle to retrieve and retain it.}
Prior studies have highlighted the importance of context understanding in AI-assisted coding~\cite{liang2024large}, and noted that a lack of contextual understanding can deter users from adopting such tools~\cite{sergeyuk2025using}. 
In our study, however, we find that the core issue is not the assistant's ability to understand context once it is available, but rather how effectively it fetches or maintains relevant context—i.e., context awareness.
As shown in Table~\ref{tab:user_feedback_categories}, reviews mentioning contextual understanding—defined as the assistant's ability to comprehend code and human instructions after the context is retrieved—are mostly positive, with 73\% explicitly highlight strong contextual understanding.
% 62\% expressing favorable sentiment. 
% Among these, 73\% explicitly highlight strong contextual understanding.
In contrast, feedback on context awareness—the assistant's ability to retrieve and maintain relevant context—is more negative, with only 38\% of related reviews are positive. 
Notably, two of the most disliked labels are related to context handling: project/codebase context support and context-memory capacity (see Table~\ref{tab:like_dislike}). 
For example, one user notes:
\textit{``[assistant] still doesn't see the class definitions in files that aren't open''} (R1, 1\ding{80}).
Others are frustrated by limited memory, such as:
\textit{``[assistant] forgets context on next question and answers irrelevantly even for simple questions''} (R22, 1\ding{80}).
We also observe that context-memory capacity frequently co-occurs with complaints about unhelpful suggestions (4 times), indicating that limited context retention can directly degrade output quality. 
For instance, one user write: 
\textit{``What AI forgets between prompts within the same thread. My code looks like Frankenstein given the amount of contradiction because it couldn't recall what I just recommended previously.''} (R79, 1\ding{80}).

\vspace*{0.1cm}
\noindent
\textbf{Finding 4. Usability matters to users—minor design issues can cause abandonment.
Users expect AI assistants to be easy to learn and use.}
While previous studies have largely focused on functionality and dependability—such as suggestion quality, privacy, and security concerns~\cite{liang2024large, sergeyuk2025using, zhou2025exploring}—usability issues have received relatively little attention. In our study, however, we find that usability is the second most discussed topic (after functionality), accounting for 12\% of all labels.
One major usability concern is onboarding difficulty. 
Although more features and stronger capabilities can enhance productivity, they also increase the learning curve. 
Among the six complaints related to onboarding, four come from extensions supporting local models, and all involve complex setup processes. 
For instance:
\textit{“Setup process is bloated. I'll wait until they make the process more streamlined.”} (R265, 1\ding{80}).
For cloud-based assistants, users report difficulty adapting to AI-driven workflows:
\textit{``While [assistant] aims to simplify coding, some users might find it challenging to adapt to the AI's suggestions and functionality, especially if they're used to traditional coding practices.''} (R240, 4\ding{80}).

Users also express frustration with suggestion interface and interaction, which are the sixth and twelfth most disliked labels in our dataset. 
Complaints include unwanted suggestion placement—e.g.,
\textit{``Annoyed suggestions show up at the top''}—and broken interaction flows, particularly related to cursor focus:
\textit{``Focus doesn't work, making chat useless...frustrated, don't use this extension.''} (R312, 1\ding{80}).
This issue refers to the assistant's ``focus'' feature, which attempts to shift user attention to the chat window. 
However, users report that this can unpredictably hijack the cursor—causing input intended for the code editor to be redirected to the chat window.\footnote{\url{https://www.reddit.com/r/Codeium/comments/1es4pdo}}
% ~\cite{codeium_focus_issue}.
Additionally, some users report severe usability breakdowns where the assistant interferes with their code:
\textit{``Messed so much with my code''} (R7, 3\ding{80}).

\vspace*{0.1cm}
\noindent
\textbf{Finding 5. Users are dissatisfied with resource consumption but generally satisfied with response time.}
Previous studies have noted that response time affects user experience, with users preferring statement-level responses within two seconds~\cite{wang2023practitioners}. 
In our study, 14 out of 17 reviews express satisfaction with response time, likely due to the increasing use of online models, which typically offer faster responses. 
Among the 32 extensions we analyzed, only four exclusively support local models.
In contrast, users frequently complain about resource consumption. 
As shown in Table~\ref{tab:user_feedback_categories}, 78\% of related reviews express dissatisfaction, making it the fourth most disliked label. 
For example:
\textit{``Uses too much resources—over 50\% CPU and more than 1 GB memory”} (R125, 1\ding{80}).
While such concerns are often associated with local models, we find that performance complaints also arise with online models, particularly when working on large files or multi-project setups.
Among the eight reviews citing high resource usage, only half refer to local models; the rest involve online models. 
One user notes:
\textit{``the extension's performance can sometimes slow down the editor, especially when working on larger files or multi projects''} (R306, 5\ding{80}).

\vspace*{0.1cm}
\noindent
\textbf{Finding 6. Users weigh pricing and feature value when adopting AI coding assistants.}
Prior studies have paid limited attention to pricing, as their data sources have primarily focused on functional performance~\cite{zhou2025exploring, liang2024large}. 
However, in our study, pricing and comparison emerged as a significant concern and plays a clear role in users' decisions to adopt or abandon AI coding assistants.
Many users express a preference for free alternatives over paid assistants. 
For example: \textit{``It's a wonderful free alternative of paid AI code assistants''} (R213, 5\ding{80}).
Users are particularly critical when an assistant lacks competitive features yet costs more than alternatives like GitHub Copilot or ChatGPT. 
One user writes: \textit{``I also tried Github Copilot, and I have to say it's a hundred levels better than [assistant], and it's cheaper to subscribe to.''} (R99, 1\ding{80}).
Some users explicitly abandon AI assistants due to pricing concerns:
\textit{``Was cool to try out but too expensive now. You are using our code to make money. So, pass for now...but I think you should have a free version (since it's using open source)''} (R42, 1\ding{80}).
This also reflects a broader ethical concern: users are dissatisfied with the practice of training on open-source code and then charging for the service. 
This issue—categorized under AI ethics label—is the second most disliked label in our study.

\vspace*{0.1cm}
\noindent
\textbf{Other Findings.}
We summarize additional insights from the remaining content in Tables~\ref{tab:user_feedback_categories} and~\ref{tab:like_dislike}, and relate them to prior work below.
The least liked subcategories—predictability, reliability, and security \& privacy—reflect high user dissatisfaction. 
Reliability and security/privacy concerns are consistent with prior studies, where users reported operational issues and fears of data leakage~\cite{zhou2025exploring, liang2024large, yang2024stealthy, yang2024unveiling}.
In contrast, task support and PL support are the second and third most liked aspects. 
Users value the assistant's help with code refactoring and test generation, aligning with Sergeyuk et al.~\cite{sergeyuk2025using}, who identified these as top tasks developers would delegate to AI assistants. 
While PL support is generally praised, some users mention a lack of support for niche languages.
The chat interface ranks fourth in user preference. Many users appreciate not having to rely on external searches~\cite{nam2024using}, though some criticize interface layout, history management, and dialog interaction.

\section{Discussion}
\label{sec:discussion}

\subsection{Implication for Improving AI Coding Assistants}
\label{subsec:implicaiton}

% In this Subsection, we give the implication based on the finding from Section~\ref{sec:results_rq2} and what users request in our analysis.

% \vspace*{0.2cm}
\noindent
\textbf{Implication 1: Improving context awareness is essential for supporting many tasks.}
As discussed in Finding 3, users complain about poor context awareness, which leads to unhelpful suggestions. 
This limitation is especially problematic for tasks like project-level refactoring, which require broader understanding of code dependencies (e.g., function changes, return types, internal logic). 
For example: \textit{``[assistant] needs to support code refactoring...changing function definitions, return types, and function contents to make sure everything compiles is a huge time sink''} (R41, 4\ding{80})
Improving repository-level context support—through codebase indexing or manual linking of relevant files—can help. 
Some assistants already support this, and users respond positively (e.g., \textit{``Indexing your repo code and adding context with @''} (R355, 5\ding{80})).
They can also enhance effectiveness by dynamic inference~\cite{sun2024when}, applying context compression techniques~\cite{ge2023context}, or fine-tuning models based on the user's repository for more personalized support~\cite{pinto2024lessons}, which can also improve the effectiveness of related tasks like program repair~\cite{luo2025unlocking, luo2025when}.

\vspace*{0.1cm}
\noindent
\textbf{Implication 2: Chat and agent interfaces need more research attention.}
While chat and agent-like capabilities are increasingly integrated into coding assistants~\cite{github_copilot_agent_2025}, prior work on understanding how assistants can help programmers has focused mainly on traditional code completion~\cite{vaithilingam2022expectation, wang2023practitioners}. 
Few studies examine the usability of chat~\cite{nam2024using, shah2025students}, and none systematically study agent interfaces. 
There is a need to understand how users interact with these modes and to design more effective interaction patterns. 
Existing work on suggestion presentation~\cite{mozannar2024show, sun2025don} can inform future research.

\vspace*{0.1cm}
\noindent
\textbf{Implication 3: Usability must be prioritized through intuitive design.}
Usability was the second most discussed topic in our study. 
Users expect intuitive design, clear documentation, and easy onboarding (e.g., \textit{``intuitive design and user-friendly interface''} (R88)). 
For example, Tabnine clearly communicates its language support levels and uses AI to provide context-aware guidance. 
As discussed earlier, improving chat and agent interactions is also crucial. 
Developers should also avoid intrusive features like ads during inference, which significantly harm user experience (R99, 1\ding{80}).
Finally, efficiency is another important aspect valued by users. 
Assistants with low resource consumption can be achieved through techniques such as model compression~\cite{shi2023compressing, shi2024greening, shi2024efficient}.

\vspace*{0.1cm}
\noindent
\textbf{Implication 4: Developers must ensure dependability to build trust.}
Perceived dependability strongly influences satisfaction~\cite{mittal1998asymmetric}. 
Users are deterred by instability, bugs, fallback to weak models, and privacy concerns~\cite{holloway2008satisfiers}. 
To build trust, assistants should be stable~\cite{yang2022natural}, reduce vulnerabilities~\cite{lyu2023chronos}, ensure fairness~\cite{lyu2025existing}, and remain transparent.
Users also request source attribution, such as:
\textit{``Display the source code link on top of the code after accepting''} (R63, 5\ding{80}), and
\textit{``Let us know where this is giving the answer from''} (R144, 5\ding{80}).
Choudhuri et al.~\cite{choudhuri2024guides} also dicuss developers' trust towards GenAI.

\vspace*{0.1cm}
\noindent
\textbf{Implication 5: Continuous user engagement and market awareness are key.}
Active communication helps assistant developers gather feedback and improve user satisfaction. 
For example, one user highlight the value of direct engagement: ``Join the Discord channel and get your questions answered, feature requests responded to, and see all the cool stuff the open source community is coming up with!'' (R340, 5\ding{80}).
Studying competitors, similar to practices in mobile app development~\cite{al2021app}, helps identify standard features and differentiation opportunities. 
Users favor assistants that offer unique capabilities beyond what others provide.
Another good example is Cursor, which actively communicates with and hears from its users via various channels.

\subsection{How much do the extensions vary in terms of capabilities?}

To better understand the diversity of AI coding assistants, we conduct a small-scale analysis by manually reviewing the descriptions and documentation of the 32 extensions analyzed in our study. 
Based on their capabilities, we categorize them into three groups: (1) extensions limited to line-level code completion (e.g., tab-to-complete); (2) extensions with chat-based functionalities, such as bug fixing or documentation generation; and (3) extensions supporting other tasks.

Among the extensions, 5 offer only line-level completion, 18 support both chat-based and line-level features, 5 offer chat capabilities alone, and 4 focus on other tasks such as documentation generation, code review, or code snippet management.
Further analysis shows that 63\% of the reviews for chat-based extensions are positive, with users focused on functionality (33.8\%) and broader discussions (18.7\% under “Other”). Line-level tools display a near-even split between likes (51\%) and dislikes (49\%), with user comments often centering on performance (12.8\%) and comparisons with other tools (12.8\%). Other extensions are the most positively received (77\% like) and are heavily praised for usability (41.9\%).

\subsection{Threats to Validity}
\label{subsec:threats}

\textbf{Internal.}
Two annotators conduct all coding and sentiment analysis. 
Although inter-rater agreement is high in the final round, their shared background may introduce confirmation bias. 
Early collaborative rounds and evolving definitions risk ``locking in'' initial interpretations. 
To mitigate this, we delay discussion until after independent blind coding. 
While seeding five top-level categories may bias consolidation toward those dimensions, annotators retain full flexibility to revise or add categories and subcategories in every round.

\textbf{External.}
Our study focuses on the VS Code Marketplace, which may raise concerns about generalizability to other development environments. 
However, VS Code is currently the most widely used IDE, with over 74\% of developers reporting it as their primary editor.
Its widespread adoption and active extension ecosystem make it a strong proxy for understanding mainstream developer interactions with AI coding assistants.
While some platform-specific differences may exist, future work could explore whether the observed patterns hold in other environments such as JetBrains IDEs or cloud-based tools.

\section{Conclusion and Future Work}
\label{sec:conclusion}

In this study, we analyze 361 reviews of 32 popular assistants. 
Using a hybrid card-sorting approach, we develop a three-level taxonomy with 8 categories, 16 subcategories, and 62 labels. 
Suggestion quality is paramount—accuracy is praised, while redundancy, incompleteness draw criticism. 
A major challenge is context awareness: assistants can interpret code but often fail to retrieve or retain relevant context. 
Usability also matters and high resource usage remains a concern. 
We also outline five implications to support the development of more effective AI coding assistants.

Future research could automate the labeling process using the taxonomy from this study and sentiment assessment to enable large-scale quantitative analysis.
A key challenge here is developing LLM-based methods that reliably detect and classify emerging labels or categories while minimizing human-in-the-loop effort, including automating the hybrid card-sorting process in iterative taxonomy refinement.

In addition, expanding data sources beyond the VS Code Marketplace to platforms like Reddit and X (formerly Twitter) could provide complementary perspectives and richer insights into user perceptions.
Another promising direction is to differentiate between types of AI coding assistants. 
For instance, future work could compare line-level code completion and chat-based assistants, and, through automated labeling, investigate how users perceive these tools differently. 
Replication package at \url{https://figshare.com/s/47c5ce42fd7e1dee69fe}.

\section*{Acknowledgments}

This research is supported by the Ministry of Education, Singapore under its Academic Research Fund Tier 3 (Award ID: MOET32020-0004). 
Any opinions, findings and conclusions or recommendations expressed in this material are those of the author(s) and do not reflect the views of the Ministry of Education, Singapore.

\balance{}

% \newpage

\bibliographystyle{IEEEtran}
\bibliography{reference}

\end{document}